\pdfoutput=1 

\documentclass[prd,aps,preprint,amsmath,nofootinbib,amssymb,eqsecnum,showkeys,tightenlines]{revtex4-1}

\usepackage{amsmath, latexsym, amssymb, hyperref, graphicx, color, multirow}
\usepackage[table]{xcolor}
\usepackage{tabularx}
\usepackage[T1]{fontenc}
\usepackage[capitalize]{cleveref}
\definecolor{nicered}{rgb}{.7,.1,.1}
\definecolor{nicegreen}{rgb}{.1,.5,.1}
\definecolor{darkblue}{rgb}{0,0,.5}
\hypersetup{colorlinks, citecolor=nicegreen,linkcolor=nicered, urlcolor=darkblue}
\usepackage{multirow}
\usepackage{slashed}
\numberwithin{equation}{section}
\usepackage{verbatim}

\begin{document}
\preprint{}

\title{Deep learning for the R-parity violating supersymmetry searches at the LHC }

\author{Jun Guo$^a$}
\email{hustgj@itp.ac.cn}

\author{Jinmian Li$^{b,c}$}
\email{jmli@kias.re.kr}

\author{Tianjun Li$^{a,d}$}
\email{tli@itp.ac.cn}

\author{Fangzhou Xu$^{a,e}$}
\email{xfz14@mails.tsinghua.edu.cn}

\author{Wenxing Zhang$^{a,d}$}
\email{zhangwenxing@itp.ac.cn}

\affiliation{$^a$CAS Key Laboratory of Theoretical Physics, Institute of Theoretical Physics, Chinese Academy of Sciences, Beijing 100190, China}
\affiliation{$^b$School of Physics, Korea Institute for Advanced Study, Seoul 130-722, Korea}
\affiliation{$^c$ Center for Theoretical Physics, College of Physical Science and Technology, Sichuan University, Chengdu, 610064, China}
\affiliation{$^d$ School of Physical Sciences, University of Chinese Academy of Sciences,
No.~19A Yuquan Road, Beijing 100049, China}
\affiliation{$^e$ Institute of Modern Physics and Center for High Energy Physics, Tsinghua University, Beijing 100084, China}

\begin{abstract}

Supersymmetry with hadronic R-parity violation in which the lightest neutralino decays into three quarks is still weakly constrained. 
This work aims to further improve the current search for this scenario by the boosted decision tree method with additional information from jet substructure. In particular, we find a deep neural network turns out to perform well in characterizing the neutralino jet substructure.
We first construct a Convolutional Neutral Network (CNN) which is capable of tagging 
the neutralino jet in any signal process by using the idea of jet image. When applied to pure jet samples, such a CNN outperforms the N-subjettiness variable by a factor of a few in tagging efficiency. 
Moreover, we find the method, which combines the CNN output and jet invariant mass,
 can perform better and is applicable to a wider range of neutralino mass than the CNN alone. Finally,
the ATLAS search for the signal of gluino pair production with subsequent decay 
$\tilde{g} \to q q \tilde{\chi}^0_1 (\to q q q)$ is recasted as an application. 
In contrast to the pure sample, the heavy contamination among jets in this complex final state renders the discriminating powers of the CNN and N-subjettiness similar. 
By analyzing the jets substructure in events which pass the ATLAS cuts with our CNN method, the exclusion limit on gluino mass can be pushed up by $\sim200$ GeV for neutralino mass $\sim 100$ GeV.

\end{abstract}

\maketitle

\section{Introduction}\label{sec:intro}

As one of the most promising new physics beyond the Standard Model (SM), 
supersymmetry (SUSY)~\cite{Nilles:1983ge, Haber:1984rc} has been copiously searched at the LHC~\cite{atlas:susy,cms:susy}. 
With the $Z_2$ R-parity~\cite{Hall_1983id}, the Lightest Supersymmetric Particle (LSP) can be 
a weakly-interacting-massive-particle dark matter (DM) candidate with 
 correct relic density~\cite{Jungman_1995df}. Moreover, the R-parity conserving (RPC) SUSY at hadron collider can be 
probed by looking for the particles with high transverse momenta and large missing energies in the final state. 
The gluino/squark masses have been excluded up to a couple of TeV~\cite{Aaboud:2017vwy,Sirunyan:2018vjp} 
at the current stage of the LHC. 

However, the R-parity is not mandatory in SUSY models. 
In contrast to the RPC scenario where the yields of colored sparticles are constrained down to $\mathcal{O}(10)$ at LHC run-II with integrated luminosity of 36 fb$^{-1}$,
some of the R-parity violating (RPV) scenarios are still weakly constrained. Thus, some improvements on 
the RPV searches are desired. 
In particular, the bounds on the RPV operators $U^c D^c D^c$, where $U^c$ and $D^c$ denote the right-handed up-type 
and down-type quark superfields respectively, are quite weak due to the large hadronic activities expected 
at the LHC~\cite{Allanach:2012vj,Durieux:2013uqa,Bhattacherjee:2013gr,Diglio:2016ynj,Buckley:2016kvr,Evans:2018scg}. In our recent work~\cite{Li:2018qxr}, the status of LHC reaches on stop and sbottom masses 
with this kind of $U^c D^c D^c$ operators are studied. We found the stop and sbottom with mass $\sim 500$ GeV are still not fully excluded. 
One of the important reasons is that the RPV scenarios were studied in the simplified model framework, such that the information of a specific signal was not fully explored.

In hadronic RPV case, the decay products of boosted heavy sparticle will be collimated, forming an single fat jet at the detector. The information from the fat jet substructure (see Refs.~\cite{Salam:2009jx,Abdesselam:2010pt,Altheimer:2012mn,Altheimer:2013yza,Adams:2015hiv,Larkoski:2017jix} for reviews) was found to be useful in improving the search sensitivities, e.g. neutralino jet substructure~\cite{Butterworth_2009qa} or stop jet substructure~\cite{Bai_2013xla,Bhattacherjee_2013tha}. 
To characterize the jet substructure, traditionally, some high-level kinematic variables such as 
mass-drop~\cite{Butterworth:2008iy} and N-subjettiness~\cite{Thaler:2010tr} are defined on the jet. 
On the other hand, all information of a jet can be inferred from the electromagnetic and hadronic calorimeters, with the basic observables being the position in the $\eta-\phi$ plane and energy deposit of each calorimeter cell. 
By identifying each cell as a pixel, and the energy deposit in the cell as the intensity (or grayscale color) of that pixel, the jet can be naturally viewed as a digital image. The recent developments of computer vision can be applied as helpful tools for us to tag the jet nature with low-level inputs. 
There are a number of works that use the jet image to discriminate hadronic $W/Z$ jet~\cite{Cogan:2014oua,deOliveira:2015xxd,Baldi:2016fql,Datta:2017rhs} and top quark jet~\cite{Almeida:2015jua,Kasieczka:2017nvn,Macaluso:2018tck} from QCD jet, and 
discriminate quark jet from gluon jet~\cite{Komiske:2016rsd,Luo:2017ncs}. 
These studies show that the jet taggers based on computer vision perform comparably or even slightly better than those based on the high-level kinematic variables. Some improved algorithms have been proposed in Refs.~\cite{Cohen:2017exh,Metodiev:2017vrx,Datta:2017lxt}.
It has been realized recently that the idea of jet image suffers from the disadvantage of low efficiency attributed to sparsity~\cite{deOliveira:2015xxd}. Machine learning techniques other than image recognition have been considered, such as using Recursive Neural Networks~\cite{Louppe:2017ipp,Cheng:2017rdo}, taking ordered sequence of jet constituents as inputs~\cite{Pearkes:2017hku}, and 
working on Lorentz vectors of jet constituents~\cite{Butter:2017cot}.

In this work, we will try to improve a realistic RPV SUSY search at the LHC by using the boosted decision tree (BDT) method~\cite{Roe:2004na} that takes into account the jet substructure information. In particular, a Convolutional Neutral Network (CNN) (for pedagogical introductions, see Ref.~\cite{Nielsen,Goodfellow-et-al-2016}) is found to be efficient in tagging the substructure of neutralino jet.
The signal process under consideration is the gluino pair production, which decays into two quarks and a neutralino. The neutralino will subsequently decay into three quarks through the hadronic RPV operator $U^c D^c D^c$. 
The main task of the CNN is to discriminate the boosted neutralino jet in this signal process from the QCD jet in SM background processes. 
Firstly, there is no prototype in the SM that is producing the same three-prong structure from three body decay as neutralino jet. Also, the mass of neutralino is an unknown parameter. We will show the change of the CNN tagging efficiency when it is applied to the neutralino mass different from the one that the CNN is trained on.  
In order to tag the neutralino jet irrespective of its production mechanism, our CNN is first trained on events of simplified process with only a visible neutralino jet in the final state.
Then it will be applied to each jet in both the signal and background events that pass all selections in the ATLAS search. 
Combining the discriminating power of the CNN scores and the jet invariant masses of leading three jets with the BDT method, the signal and background can be separated further, leading to a better search sensitivity.

The paper is organized as follows. In Sec.~\ref{sec:na}, the architecture of 
the CNN that is adopted in this paper will be given. Sec.~\ref{sec:train} discusses the training process and performance of the CNN on an simplified signal process. Its application to a realistic RPV gluino search is studied in Sec.~\ref{sec:lhc}. Our conclusion is provided in Sec.~\ref{sec:conc}.

\section{The CNN Architecture} \label{sec:na}

\begin{figure}[htb]
\begin{center}
\includegraphics[width=0.7\textwidth]{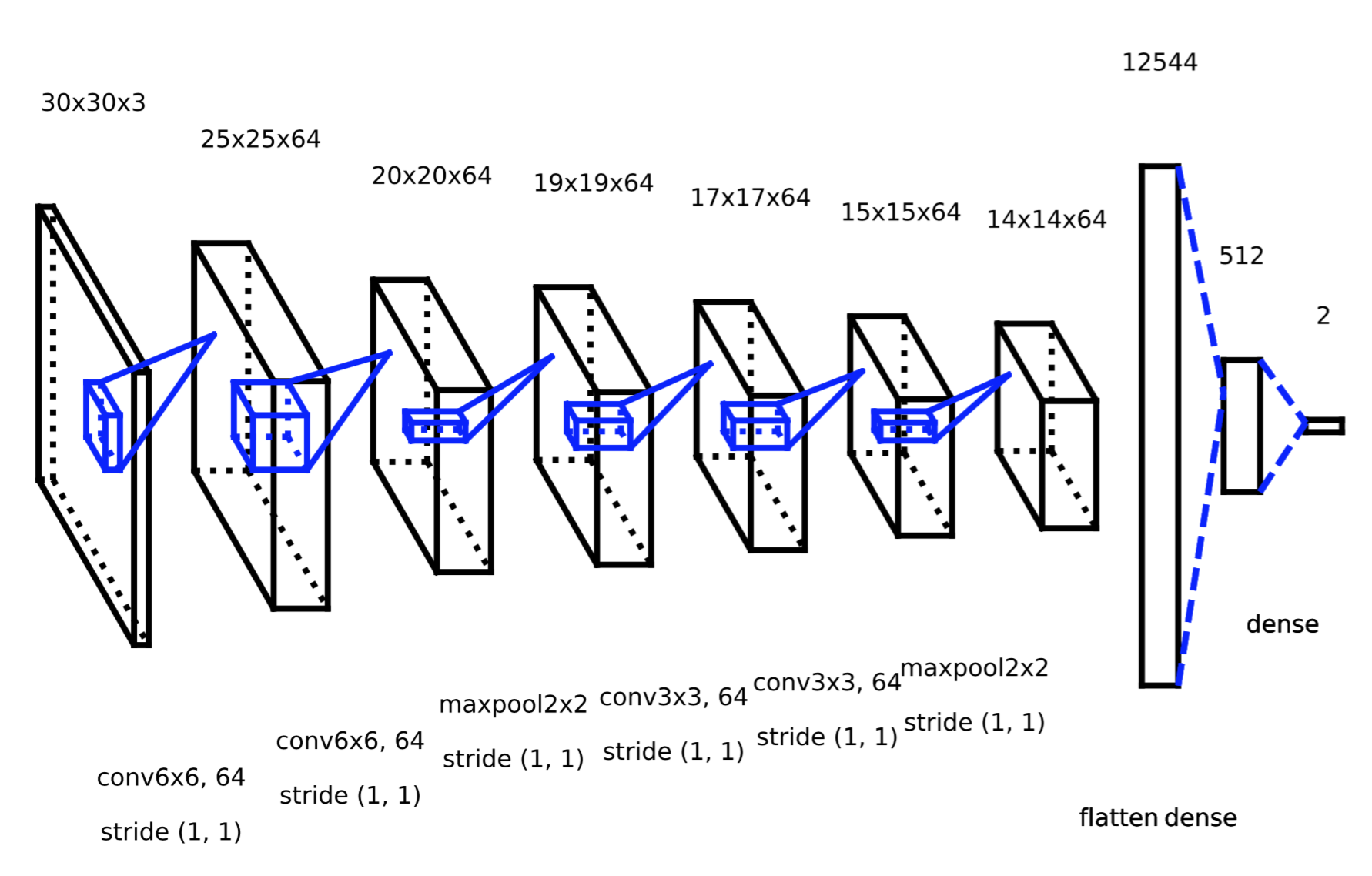}
\end{center}
\caption{\label{fig:arch} The architecture of our CNN for one of the parameter choices. }
\end{figure}

There exist many CNN architectures, such as the \texttt{VGGNet}~\cite{DBLP:journals/corr/SimonyanZ14a} and \texttt{ResNet}~\cite{DBLP:journals/corr/HeZRS15}. They have been proved to be very successful in classifying images of either large size (in PASCAL Visual Object Classes dataset~\cite{Everingham10}) or small size (in \texttt{CIFAR-10}~\cite{KAlex} dataset). 
As for our case, due to the limited angular resolution of the detectors at hadron collider, the jet image is usually smaller than $30 \times 30$ pixels. It has similar size to images in \texttt{CIFAR-10}~\cite{KAlex} dataset. 
Inspired by the \texttt{VGGNet} architectures that was optimized for \texttt{CIFAR-10} dataset, the sketch of our CNN architecture is shown in Fig.~\ref{fig:arch}. 

The input consists of three layers defined as the energy distribution of all particles, the energy distribution of charged particles and the number of charged particles in calorimeter cells. The more detailed jet image preprocessing will be introduced later. 
Those data are then passed through two iterations of two convolutional layers with ReLU activation and a max-pooling layer. The size and the total number of the convolution kernel (also called filter) in each convolutional layer are free parameters. In practice, we need trial and error to figure out the best choice. In the figure, at the first step of iteration, the input is convoluted twice by 64 filters with same size of $6\times 6$, followed by max-pooling with filter of size $2\times 2$ and with stride of unit. While at the second step of iteration, the size of convolutional filters are reduced to $3\times 3$. The total number of filters in each convolutional layer and the filter in pooling layer remain the same as the first iteration. 
The feature map is flattened and read by the fully connected neural network (FCNN). There are 512 neural nodes defined in the hidden layer of the FCNN where the ReLU activation function has been adopted. The final output layer contains two nodes with sigmoid activation function. With the output value of each node between [0,1], it can be used to characterize the probability of being either signal or background.



\section{Training and Testing of CNN} \label{sec:train} 

Our goal is to employ the CNN that can recognize the jet image of neutralino from jet images of quark and gluon, so that the signal processes with neutralino in the final state can be separated further from backgrounds. To make our CNN a general neutralino jet recognizer which is not specific to any detailed production process, the training of CNN is based on the signal event samples with only one visible neutralino in the final state which subsequently decay into three quarks. 
Throughout the work, the hard-scattering signal and background events as well as the neutralino decay are simulated by the MadGraph5\_aMC@NLO program~\cite{Alwall:2014hca}. Pythia8 package~\cite{Sjostrand:2006za} is used to perform the parton shower and hadronization. The detector effects are simulated by the Delphes3~\cite{deFavereau:2013fsa} with ATLAS configuration card, in which the jet reclustering algorithm is implemented via the FastJet~\cite{Cacciari:2011ma} software. 
Our CNN is implemented in Python, using the deep learning library Keras~\cite{keras}. 
 
The training and testing samples are generated and processed as follows. Firstly, the signal events with single visible neutralino jet are generated by $p p \to \tilde{\chi}^0_1 \tilde{\chi}^0_2$ process in SUSY model, with $\tilde{\chi}^0_2 \to b c s$ through the $U^c_2 D^c_2 D^c_3$ operator~\cite{Li:2018qxr}. The $\tilde{\chi}^0_1$ is assumed to be stable here which leaves nothing inside the detector~\footnote{This is a trick in generating process independent neutralino jets for training and testing. In the next section, considering a complete model, $\tilde{\chi}^0_1$ is the LSP that decays into three quarks, 
{\it i.e.}, $ b c s$.}. As a benchmark, we choose the mass of $\tilde{\chi}^0_2$ to be 100 GeV. Its transverse momentum is required to be $p_T(\tilde{\chi}^0_2)>200$ GeV so its decay products are collimated and behave as a jet at detector. 
Furthermore, it is obvious that the neutralino jet image will be varying if the polar angle (or pseudorapidity) of the neutralino is changed. To consider this effect, two classes of signal events sample are generated: one with requirement of $|\eta(\tilde{\chi}^0_2)|<0.1$ (central sample) and the other allows much larger pseudorapidity $|\eta(\tilde{\chi}^0_2)|<2.5$ (wide sample). 
Secondly, the background events in training and testing are generated by $p p \to j \tilde{\chi}^0_1 \tilde{\chi}^0_1$ in SUSY model, where $j$ can be either quark or gluon and $\tilde{\chi}^0_1$ is stable at detector. As in signal event generation, the transverse momentum of $j$ is required to be $p_T(j) >200$ GeV and two classes of background samples with cuts of $|\eta(\tilde{\chi}^0_2)|<0.1$ and $|\eta(\tilde{\chi}^0_2)|<2.5$ are defined. 
It should be noted that during the training and testing stage, the initial state radiation and multi-particle interaction have been turned off in Pythia8 for both signal and background event generation~\footnote{These effects will be included when considering a realistic gluino search in the next section.}. Thus, their contaminations to the target jet image are suppressed and the CNN can grab the important features of target jets more efficiently.
Thirdly, in both signal and background events, jets are reconstructed by the anti-$k_t$ algorithm~\cite{Cacciari:2008gp} with cone size $R=1.0$. The minimal transverse momentum of target jet should be 100 GeV~\footnote{This requirement is looser than that at parton level because we find the reconstructed jet can be softer than the parton level jet sometimes due to large angle splitting.}. An event will be dropped if there is no jet with $p_T(j)>100$ GeV. In case of more than one jet with $p_T(j)>100$ GeV in an event, the jet with highest $p_T$ is chosen. For signal events, we also require the selected jet lie within a cone size of $R<1.0$ to the parton level $\tilde{\chi}^0_2$. 

At this stage, each event has been associated with a single jet, which is expected to be neutralino jet (QCD jet) for signal (background) event. Next, we need to convert the jet information into grid image. Given a jet, its hardest constituent is located on the $\eta-\phi$ plane. Afterwards, a grid with step of $0.1 \times 0.1$ and size of $30\times 30$,
which is centralized at the hardest constituent, is defined. 
Based on the grid and jet constituent information, we can define three different layers for jet image: (1) The layer that shows the energy grid of all jet constituents, where the energies of jet constituents belong to the same cell are added up; (2) As in the first layer, but only the energy of charged jet constituents are taken into account; (3) The layer counts the number of charged jet constituents in each cell. 
Since the CNN is found to be most efficient in dealing with numbers between [0,1], all numbers in each layers are divided by the maximum value in that layer, e.g. the maximum energy of the cell in the first layer. 
We will not apply any more image preprocessing procedures, such as rotation and flipping, because they were found to decrease the performance of our CNN (same finding as in Ref.~\cite{Kasieczka:2017nvn}). 

Finally, to use our data set in a more efficient way (we have generated one million signal and background events for training), 30 epochs are required during the training process. And to avoid the over-training problem, an independent dataset of one million signal and background events is used for testing.  

\begin{figure}[htb]
\begin{center}
\includegraphics[width=0.45\textwidth]{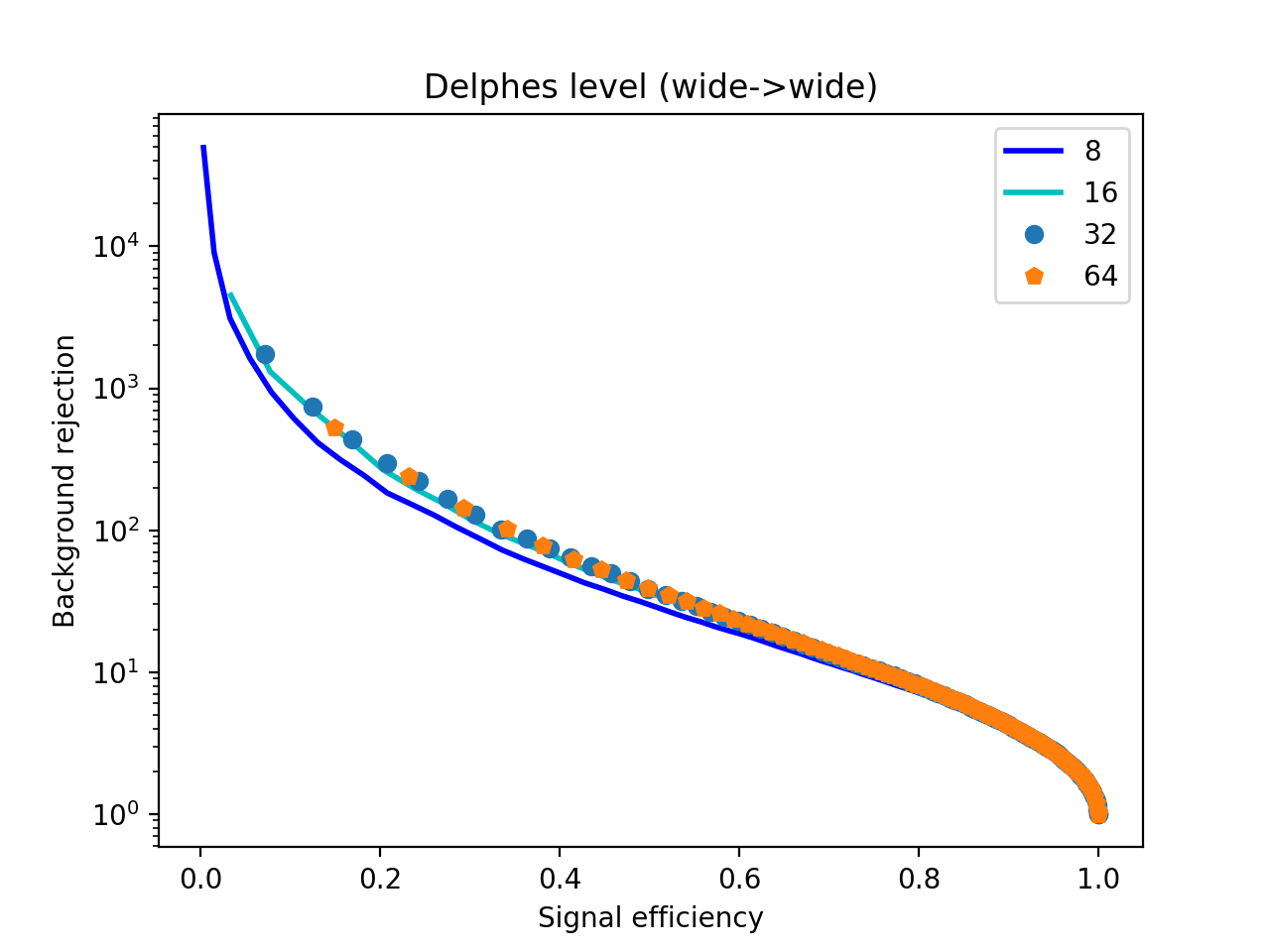}
\includegraphics[width=0.45\textwidth]{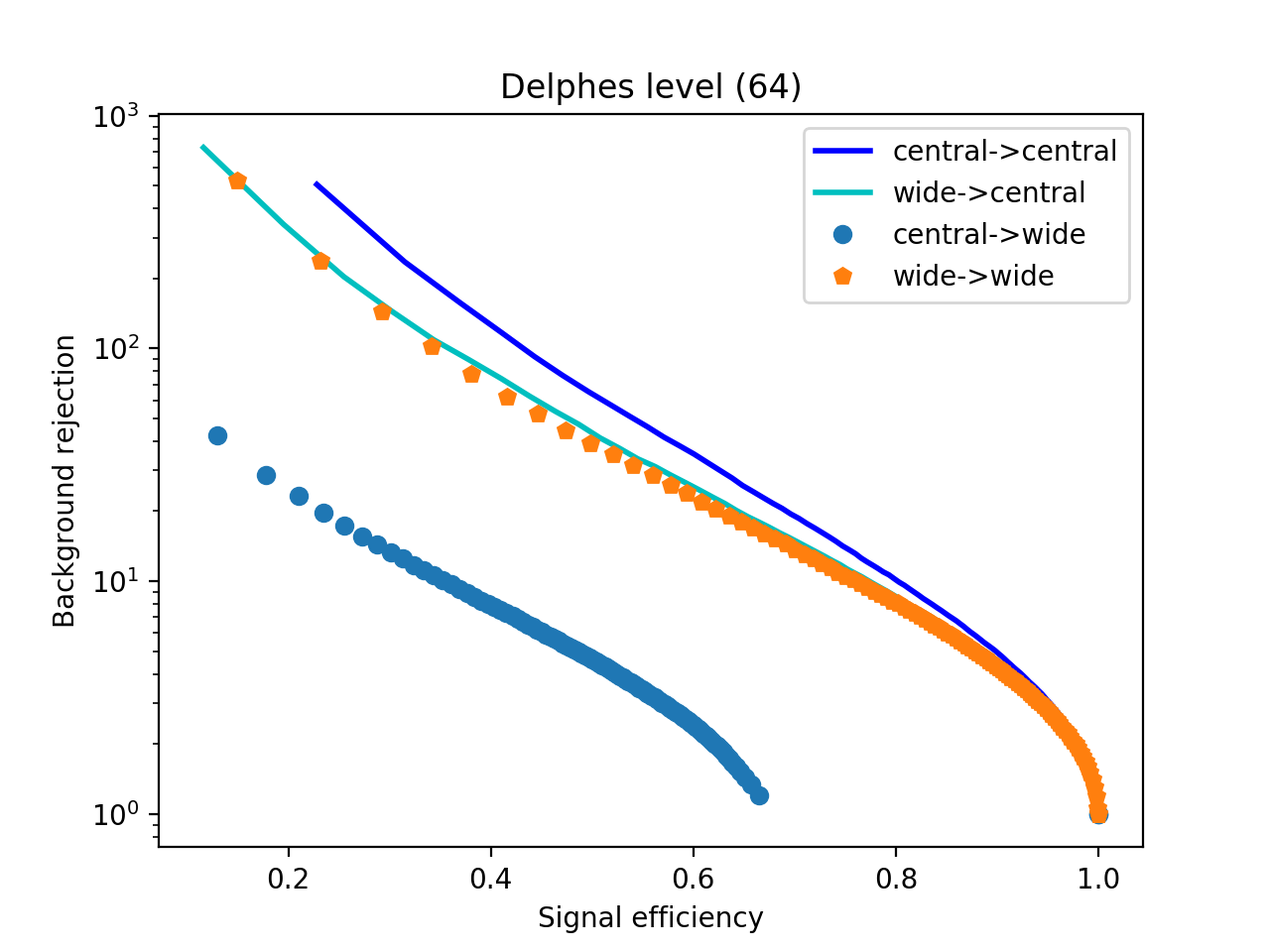}
\end{center}
\caption{\label{fig:param} Left panel: performances of the CNNs with different number of convolutional filters in convolutional layers. Right panel: performances of the CNNs that have been trained on central sample (wide sample) and applied to either central sample or wide sample. Details of other parameter choices are discussed in the text.}
\end{figure}

There are a number of free parameters in the CNN that can only be optimized through trial and error, including sizes and numbers of convolutional kernels in the convolutional layers, the dropout rates after two iterations and FCNN, the number of nodes in the hidden layer of the FCNN and the learning rate in NAdam algorithm~\cite{DBLP:journals/corr/KingmaB14}. We find the performance of the CNN only mildly depends on these parameters. In the left panel of Fig.~\ref{fig:param}, the performances of the CNNs with the number of convolutional filter in convolutional layers being 8, 16, 32 and 64 are shown (same number is adopted in all convolutional layers). The CNN with convolutional kernels more than 16 performs equally well, they are slightly better than the one with 8 convolutional kernels. 
To obtain the results, we have taken the size of the convolutional kernel to be $6 \times 6$~\footnote{We find the CNNs with filter sizes of $2\times2$ and $4\times 4$ perform worse.}, the dropout rate in two iterations as 0.25 while it is 0.5 for the FCNN. The number of nodes in the hidden layer is 512 and the learning rate is taken to be 0.001. This parameter choice will be used throughout this work. Even though the number of trainable parameter here ($\sim 6.5$ million) is larger than the size of the training sample, our CNN is still working fine because of the following two reasons. First, we have tried the CNN with much smaller parameter set (with 8 filters in convolutional layers and 64 nodes in the hidden layer of the FCNN, the parameter number is $\sim 0.1$ million), its performance is slightly worse than the one shown in Fig.~\ref{fig:arch}. Second, the trained CNN has been tested on an independent event sample, which gives similar accuracy. So the CNN is not overtrained on the training sample.
Note that we have defined two CNNs that are trained and tested on central sample and wide sample of signal and background events, respectively. The results presented in the left panel correspond to the wide sample trained CNN applying to another independent wide sample. 
In the right panel, to characterize the dependence of the jet image feature on the jet pseudorapidity, we show the performance of these two sets of the CNNs (both with 64 filters in all convolutional layers) on different samples. 
There is no doubt that the central jet ($|\eta|<0.1$) is easier to tag than the jet within wide pseudorapidity range ($|\eta|<2.5$). The CNN trained and tested on the central sample is not working for tagging neutralino jet in the wide sample, 
mainly because the features captured by the CNN in central sample are not useful for wide sample. On the other hand, the CNN trained and tested on the wide sample performs well in tagging neutralino jet in the central sample, even though it is slightly worse than the CNN that is trained and tested directly on the central sample. 
This means we do not have to limit our analysis to the phase space with target jet in the central region.  
It is especially useful in a realistic signal search at the LHC, so that more signal events can be saved. 
In the following, we will keep using the CNN that is trained and tested on the wide sample with filter number in each convolutional layer being 64.

\begin{figure}[htb]
\begin{center}
\includegraphics[width=0.6\textwidth]{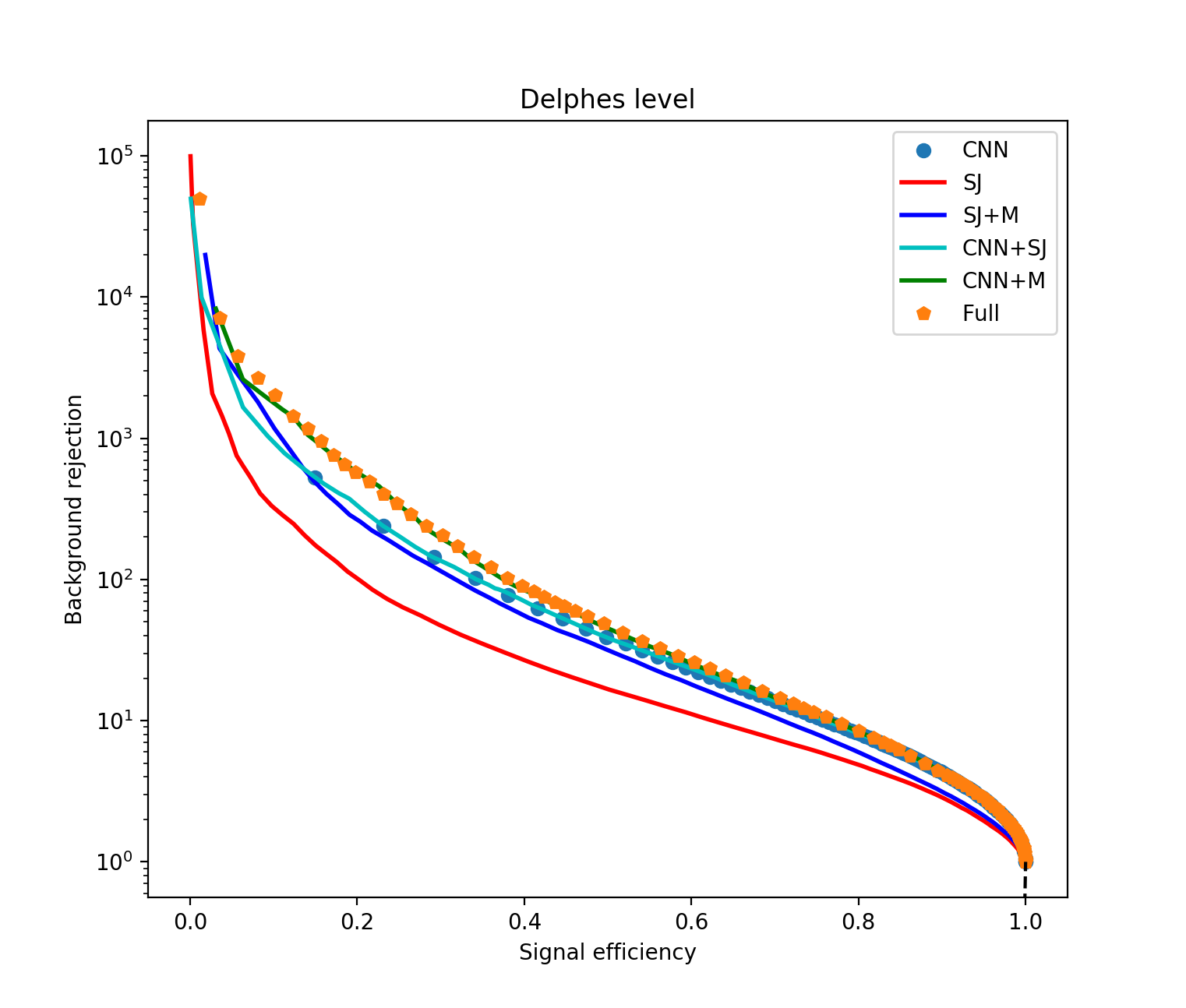}
\end{center}
\caption{\label{fig:comp} Comparison among the performances of different methods in discriminating the neutralino jet 
from QCD jet, SJ denotes the N-subjettiness variable, M is the jet invariant mass, and 
the combination of different variables are managed by the BDT method. }
\end{figure}

We should compare the performance of our CNN with those high-level jet substructure variables. Among these, the N-subjettiness is a general and effective discriminating variable that can characterize the multi-prong structure of a jet. It is defined as~\cite{Thaler:2010tr} 
\begin{align}
\tau_N = \frac{\sum_k \min \{  \Delta R_{1,k},  \Delta R_{2,k}, ... , \Delta R_{N,k} \} }{\sum_k p_{T,k} R_0 }~,
\end{align}
where $k$ runs over all constituent particles in a given jet, $p_{T,k}$ is transverse momentum of the $k$th constituent, $R_{J,k}$ is the distance between a candidate subjet $J$ and the $k$th constituent in the $\eta-\phi$ plane, and $R_0$ is the characteristic jet radius that is used in the original jet clustering algorithm. 
A jet with N-prong will have $\tau_N \sim 0$ when all of its constituents are aligned with candidate subjets while $\tau_I \gg 0$ for $I <N$ because there are constituents distributed away from the candidate subjet directions. As a result, the variable $\tau_N/\tau_{N-1}$ is found to be efficient in tagging jet with N-prong structure. In our case, the neutralino jet substructure can be tagged by $\tau_3/\tau_2$. 
The performance of the N-subjettiness technique is shown by the red solid line in Fig.~\ref{fig:comp}. We find that the performance of our CNN (represented by blue dots) is a few times better than that of the N-subjettiness.  
Moreover, the jet invariant mass is a powerful discriminating variable that is independent of N-subjettiness. To combine the discriminating power of both variables, the BDT method is adopted. 
Because the BDT only needs to learn two dimensional information here, a relatively small size of forest should be enough. It uses a 100 tree ensemble that requires a minimum training events in each leaf node of 2.5\% and a maximum tree depth of three.  The rest of the parameters are set to default ones in the TMVA package~\cite{Hocker:2007ht}. It is trained on half of the reconstructed neutralino and QCD jets and is tested on the rest of the jets ($\sim $0.5 million each). To avoid overtraining, the Kolmogorov-Smirnov test~\cite{Chakravarty:109749} in the BDT training and testing is required to be greater than 0.01~\footnote{In practise, we find the Kolmogorov-Smirnov tests are always greater than 0.1 for both neutralino and QCD jets.}.
The performance of the combination of N-subjettiness and jet invariant mass is given by the blue solid line, which shows the similar tagging efficiency as the CNN alone. 

Meanwhile, it is worth finding out whether our CNN is clever enough to learn both the N-prong structure and the jet invariant mass~\cite{Chang:2017kvc,Datta:2017rhs}. This can be seen through the tagging efficiencies of their combinations. In Fig.~\ref{fig:comp}, the performances of the CNN + N-subjettiness (SJ) and the CNN +  jet invariant mass (M) are shown by cyan and green solid lines, respectively. The combination of their sensitivities are managed by the BDT method, with the same parameters as introduced above. The CNN+SJ does not show much more improvement than the CNN alone. While the tagging efficiency can be improved by a factor of a few after including the jet invariant mass. 
Thus, we can conclude that the full information of prong structure in a jet can be learned by the CNN but the jet invariant mass cannot be directly extracted from the jet image by current method. One of the reasons is that the image preprocessing procedures do not respect the Lorentz symmetry, so the jet invariant mass is broken down in the preprocessing~\cite{deOliveira:2015xxd,Larkoski:2017jix,Kasieczka:2017nvn}. 

\begin{figure}[htb]
\begin{center}
\includegraphics[width=0.45\textwidth]{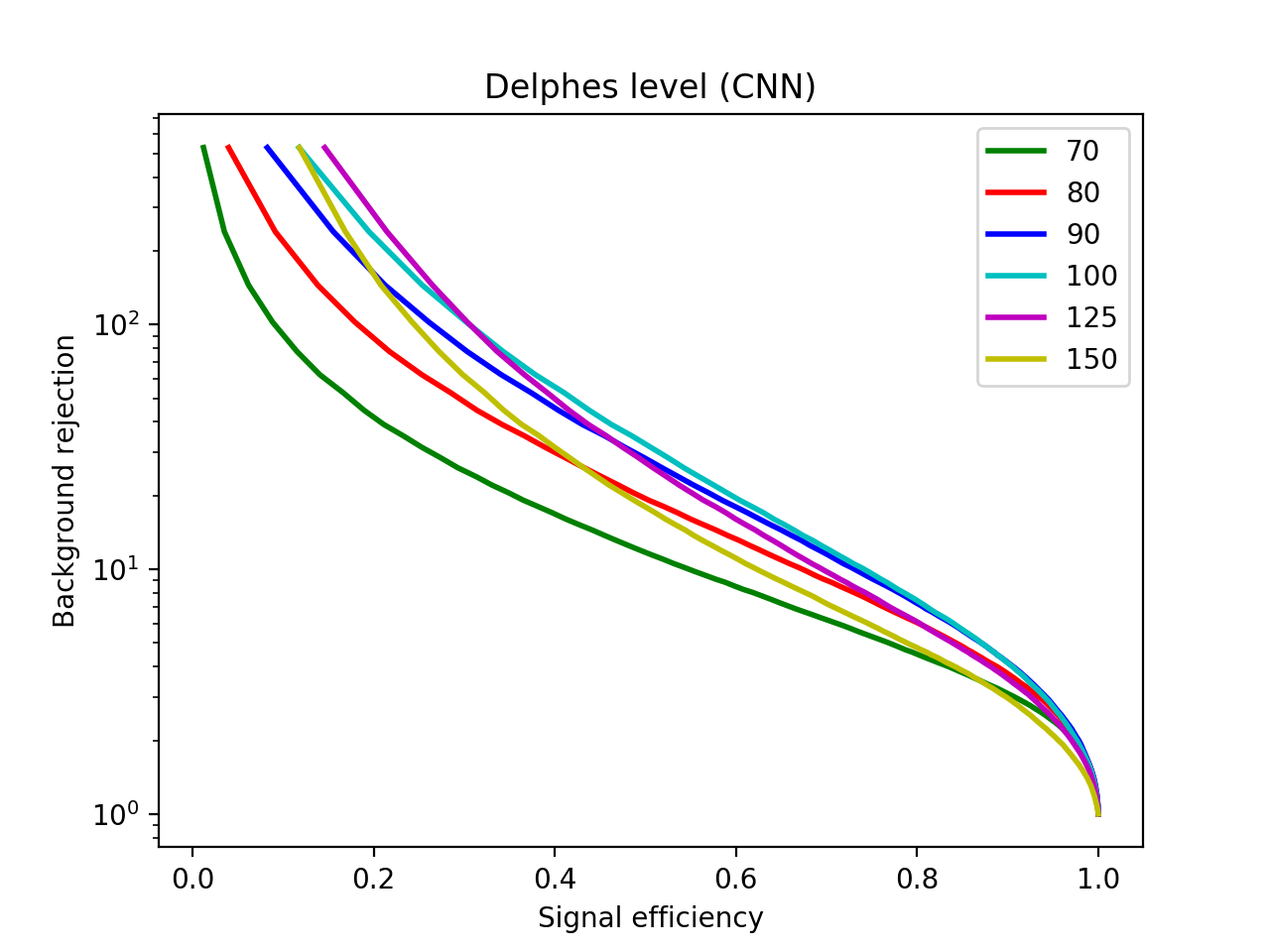} 
\includegraphics[width=0.45\textwidth]{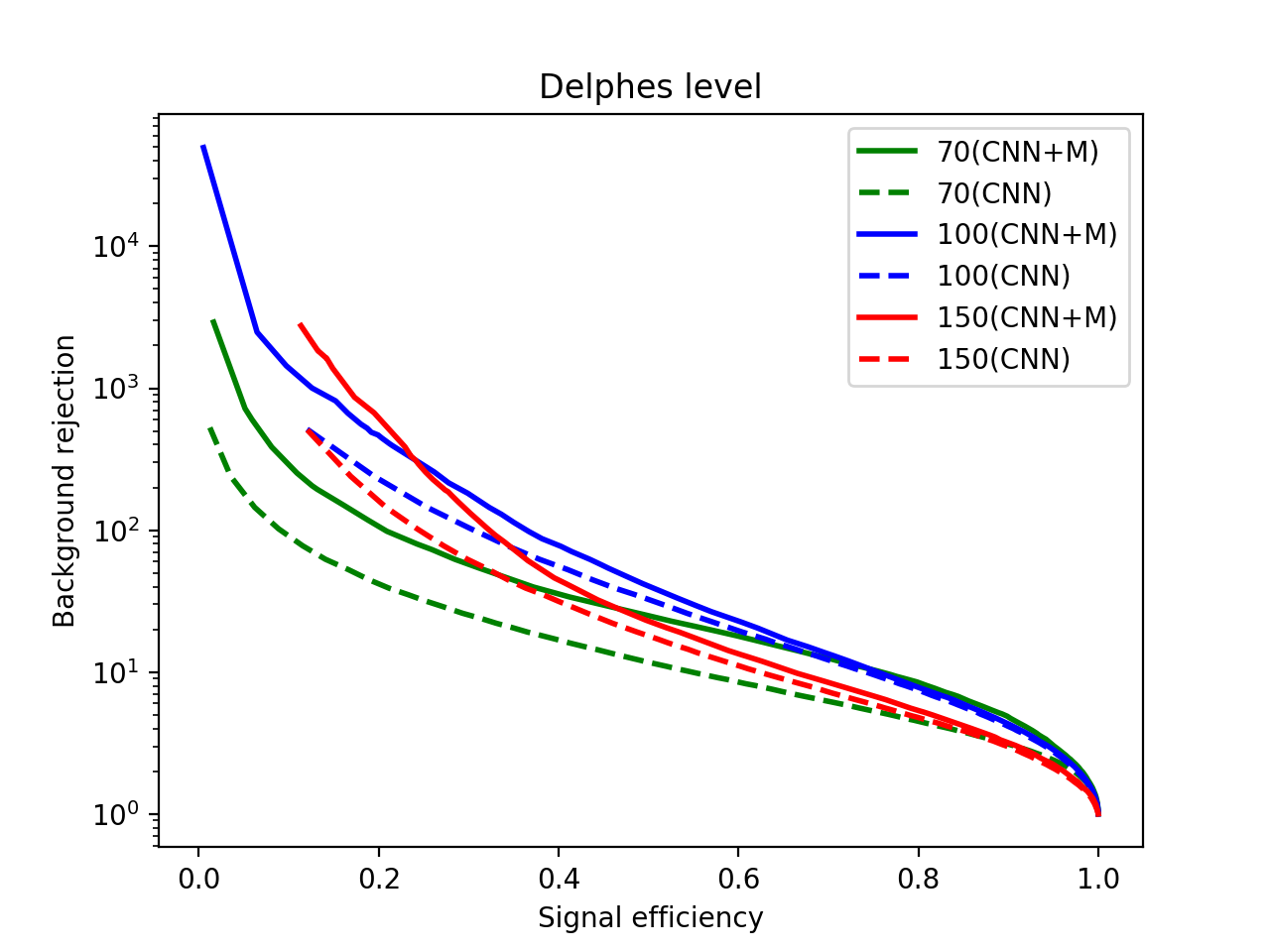}
\end{center}
\caption{\label{fig:masses} Left panel: performances of the CNN when applied to the samples with different neutralino masses.  Right panel: performances of the method that combine the information of the CNN and jet invariant mass. In both panel, the CNN is trained with $m_{\tilde{\chi}^0_2} =100$ GeV event sample only.  The numbers in the legend indicate the neutralino masses of event samples. } 
\end{figure}

In the above study, the neutralino mass has been taken to be 100 GeV in all event samples. 
In practice, especially our method is trying to improve the signal discovery sensitivity, the neutralino mass is an unknown parameter. 
It will be unrealistic to have the CNNs with the same neutralino mass as the signals that we want to probe. 
One way~\footnote{One can also train a neural network on the event sample that is containing events of all neutralino masses as in Ref.~\cite{Baldi:2016fzo}.} in discovery is to train several CNNs, each at a chosen neutralino mass, and apply those CNNs to a wide range of neutralino mass. Then, for any given neutralino mass, the CNN, 
which was trained on the closest neutralino mass, is able to tag the signal efficiently. 
The generality of the CNN, which is trained on a fixed neutralino mass, can be seen in Fig.~\ref{fig:masses}. 
In the left panel, we show the performances of the CNN on event samples with neutralino mass in the range of [70,150] GeV, where the CNN is trained with $m_{\tilde{\chi}^0_2} =100$ GeV event sample only.
We find that the neutralino mass varying in the range of [90,125] GeV does not reduce the sensitivity much and the CNN is more vulnerable to lower neutralino mass. 
On the other hand, the CNN can be more useful if it is used in combination with jet invariant mass (CNN+M). 
In the right panel, the performances of the combinational CNN+M on different neutralino masses are shown. 
The information from jet invariant mass helps improving the tagging efficiency a lot, especially in the light neutralino mass region, and compensating for the weakness of the CNN. The efficiency of the CNN+M method only mildly depends on the neutralino mass. 
To conclude, including jet invariant mass can not only improve the tagging efficiency but also extend the application of our CNN. 
They should be used together in realistic signal searches.

\section{Application to the LHC gluino search} \label{sec:lhc}

Having shown the power and generality of our CNN method, we are ready to show its explicit application in a RPV gluino search~\footnote{An attempt to improve the same search using the whole event image with the CNN was studied in Ref.~\cite{Bhimji:2017qvb}. }. The signal process is the gluino pair production, in which each gluino decays into two quarks and a neutralino. 
The neutralino decays through the hadronic RPV operator into three quarks. 
This signal has been searched by the ATLAS Collaboration in Ref.~\cite{ATLAS-CONF-2016-057}. For neutralino with mass $\sim 100$ GeV, gluino lighter than $\sim 1.1$ TeV has been excluded.  The dominant background process in the search is the QCD multi-jet background. 
In this Section, we will show how the CNN helps to improve the ATLAS gluino search. Before that, we need to recast the experimental analysis on both signal and background. 

The QCD multi-jet process is simulated by the MadGraph5 framework at the leading order~\footnote{The higher order QCD corrections, which change the distributions of jet multiplicity, jet pseudorapidity and jet transverse momentum, can only have indirect influences on the jet substructure, such as more contaminations between jets due to higher jet multiplicity, spread jet profile for larger pseudorapidity and/or smaller transverse momentum. Our results are insensitive to these effects because the parton showing with Pythia8 includes all the leading logarithmic contributions and our CNN is capable to tag jet in a wide range of pseudorapidity and transverse momentum.}. According to the cuts adopted in the ATLAS analysis, we only consider the multi-jet processes with 4/5 jets at parton level, each jet should have $p_T > 200$ GeV and $|\eta|<2.0$. And the matching of these processes are handled by MLM method~\cite{Mangano:2006rw} in the MadGraph5. Events with higher jet multiplicity are obtained after performing the initial state radiation and final state radiation in Pythia8. 
The signal events are generated at the leading order as well, based on the benchmark points that have neutralino mass in [50,200] GeV and gluino mass in the range of [1,2] TeV with step size 50 GeV. 

We recast the ATLAS analysis~\cite{ATLAS-CONF-2016-057} as follows. (1) For each event, large-$R$ jets are reconstructed by anti-$k_T$ algorithm with radius parameter $R=1.0$. A "trimming" process~\cite{Krohn:2009th} with subjet radius parameter of $R_{\text{subjet}} =0.2$ and the minimal transverse momentum fraction of 5\% is applied on each large-$R$ jet. 
The resulting trimmed large-$R$ jets are required to have $p_T>200$ GeV and $|\eta|<2.0$. 
The analysis only selects the events with at least four trimmed large-$R$ jets ($N_{\text{jet}} \geq 4$) in which the leading one should have $p_T>440$ GeV. (2) Meanwhile, the small-$R$ jets of each events are reconstructed by anti-$k_T$ algorithm with radius parameter $R=0.4$. They are required to have $p_T>50$ GeV and $|\eta|<2.5$. These jets are used to count the number of $b$-tagged jets ($N_b$) in the final state. The $b$-tagging efficiency is taken to be 70\%~\cite{ATL-PHYS-PUB-2016-012} with mis-tagging rates for the charm- and light-flavor jets of 0.15 and 0.008, respectively. (3) Two discriminative variables are defined for each event: the total jet mass variable ($M^{\Sigma}_J$)~\cite{Hook:2012fd,Hedri:2013pvl,Cohen:2014epa,Aad:2015lea} which is the scalar sum of invariant masses of four leading trimmed large-$R$ jets and the pseudorapidity difference between the two leading trimmed large-$R$ jets ($|\Delta \eta_{12}|$). (4) Four signal regions are defined in Tab.~\ref{tab:sr}.

\begin{table}[htb]
\begin{center}
\begin{tabular}{|c||c|c|c|c|c|c|} \hline
Signal region & $N_{\text{jet}}$ & $N_b$ & $M^{\Sigma}_J$ & $|\Delta \eta_{12}|$ & observed   & SM predicted \\ \hline
4jSR &  $\geq 4$ & - & $>0.8$ TeV & $<1.4$  & 122 & $151\pm 15 \pm 17 \pm 20$ \\
4jSRb1 & $\geq 4$ & $>0$  & $>0.8$ TeV & $<1.4$ & 46 & $61\pm 10\pm 6\pm 12$  \\
5jSR & $\geq 5$ & -  &  $>0.6$ TeV  & $<1.4$  & 64 & $51.4 \pm 7.7 \pm 7.2 \pm 6.5$   \\
5jSRb1 &   $\geq 5$  &  $>0$  &  $>0.6$ TeV &  $<1.4$ & 30 & $18.2 \pm 4.2 \pm 2.5 \pm 3.0$ \\\hline
\end{tabular}
\caption{\label{tab:sr} The definitions, the expected numbers of background events, 
and the observed event numbers of four signal regions in the ATLAS analysis~\cite{ATLAS-CONF-2016-057}.  Three components of background prediction uncertainty in the seventh column are  statistical uncertainty, residual $p_T$-dependence uncertainty, and the Monte Carlo-based non-closure uncertainty, respectively. }
\end{center}
\end{table}

Because we are interested in the low neutralino mass region, the 4jSRb1 signal region provides the most sensitive probe. 
Only the signal and background events, which can pass all of the selections of 4jSRb1 signal region,
 are kept for later analysis. 
In the simulation, the selected signal and background event numbers are guaranteed to be around 10K, to suppress the statistical uncertainty.
The cross section for signal at this stage can be calculated as the $\sigma^{13}(\tilde{g}\tilde{g}) \times \epsilon^{\text{4jSRb1}}$, where $\sigma^{13}(\tilde{g}\tilde{g})$ is the gluino pair production cross section at the 13 TeV LHC which can be calculated 
at the next-to-leading-order by Prospino2~\cite{Beenakker:1996ed} and $\epsilon^{\text{4jSRb1}}$ is the selection efficiency of the 4jSRb1 signal region that is obtained from our recasted analysis. The background cross section ($\sigma^{\text{BG}}$) at this stage is simply estimated by the numbers in the "SM predicted" column of Tab.~\ref{tab:sr} divided by the integrated luminosity of the analysis $\mathcal{L}=14.8$ fb$^{-1}$.

\begin{figure}[htb]
\begin{center}
\includegraphics[width=0.3\textwidth]{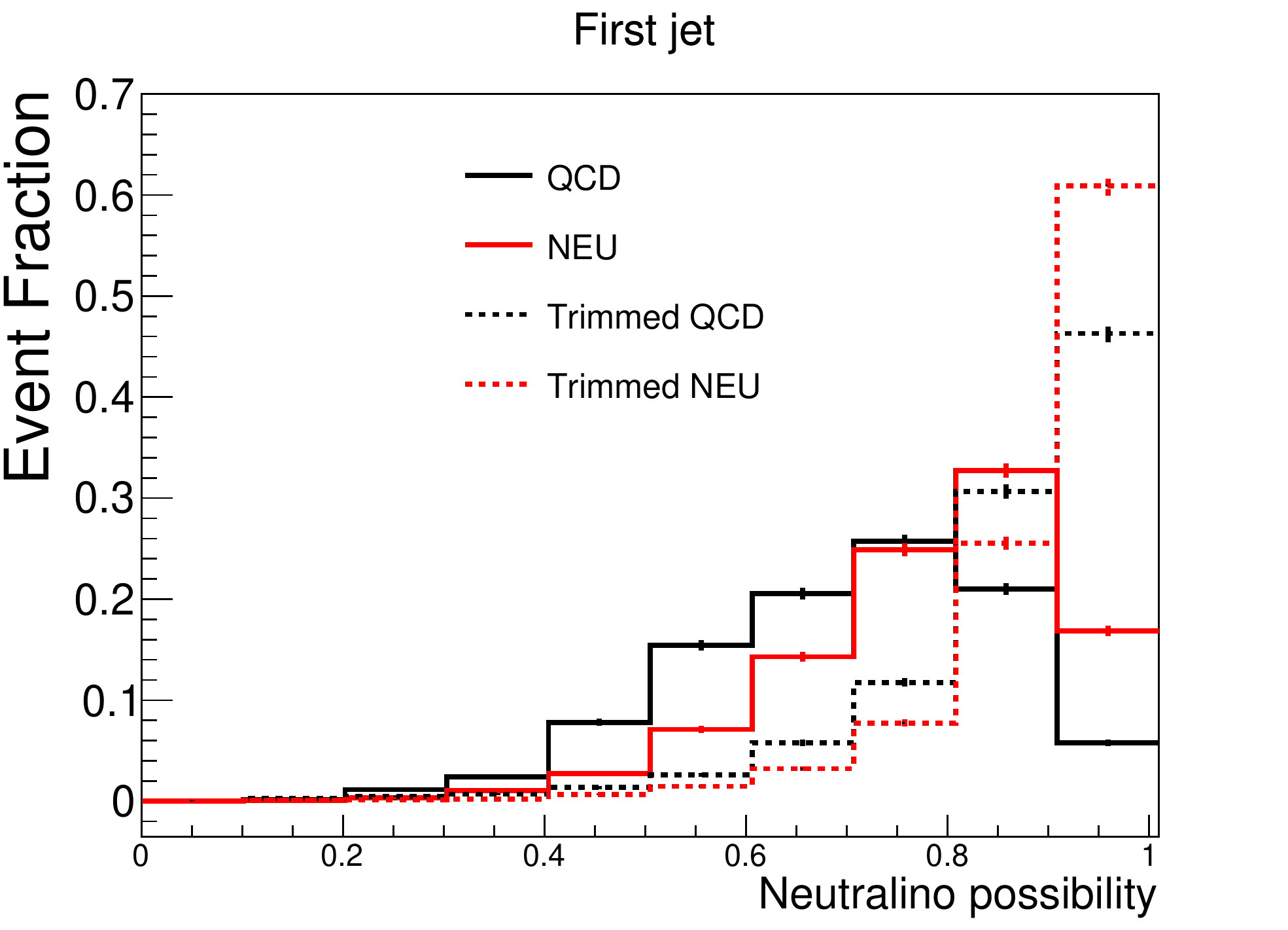} 
\includegraphics[width=0.3\textwidth]{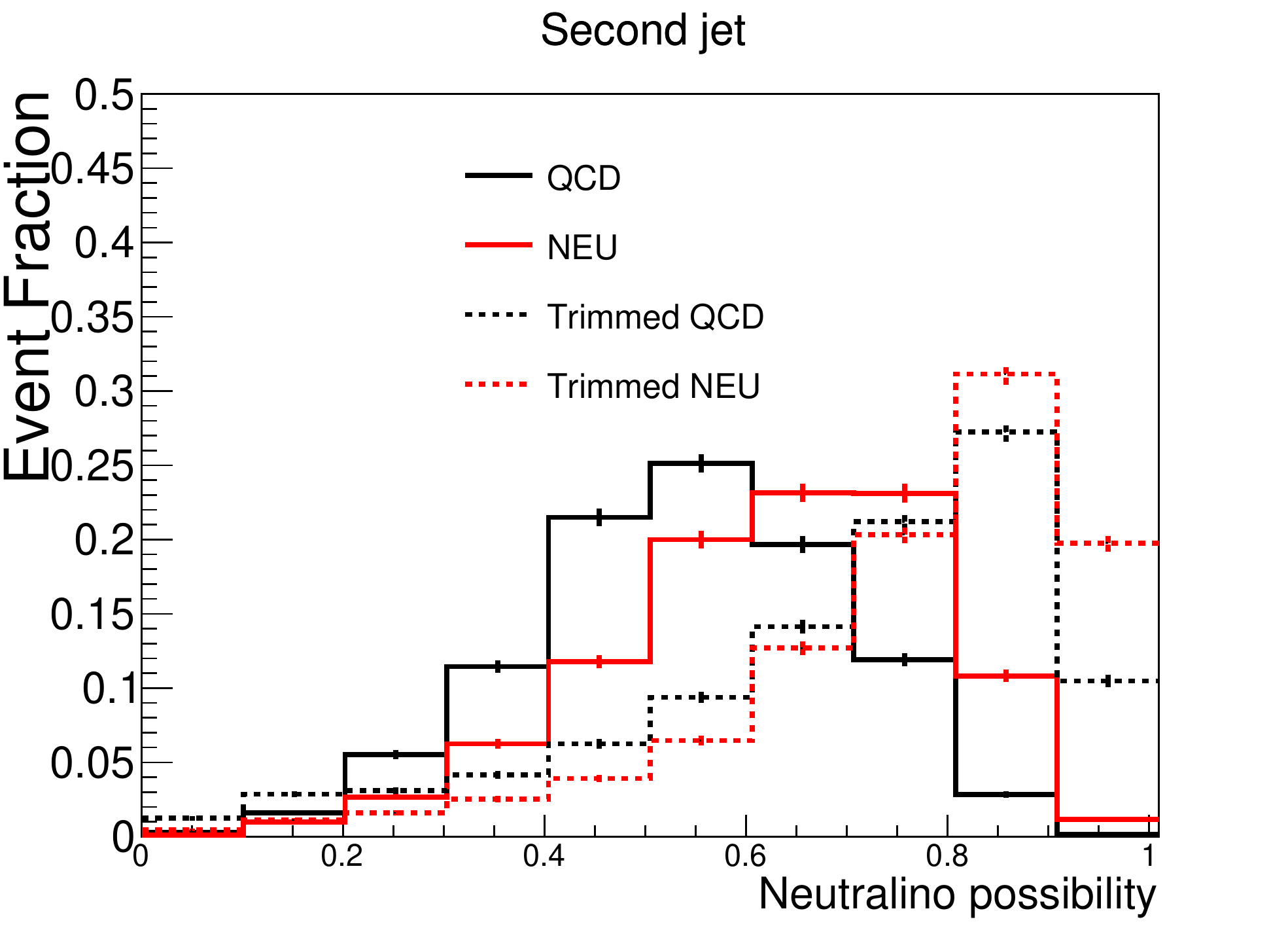}
\includegraphics[width=0.3\textwidth]{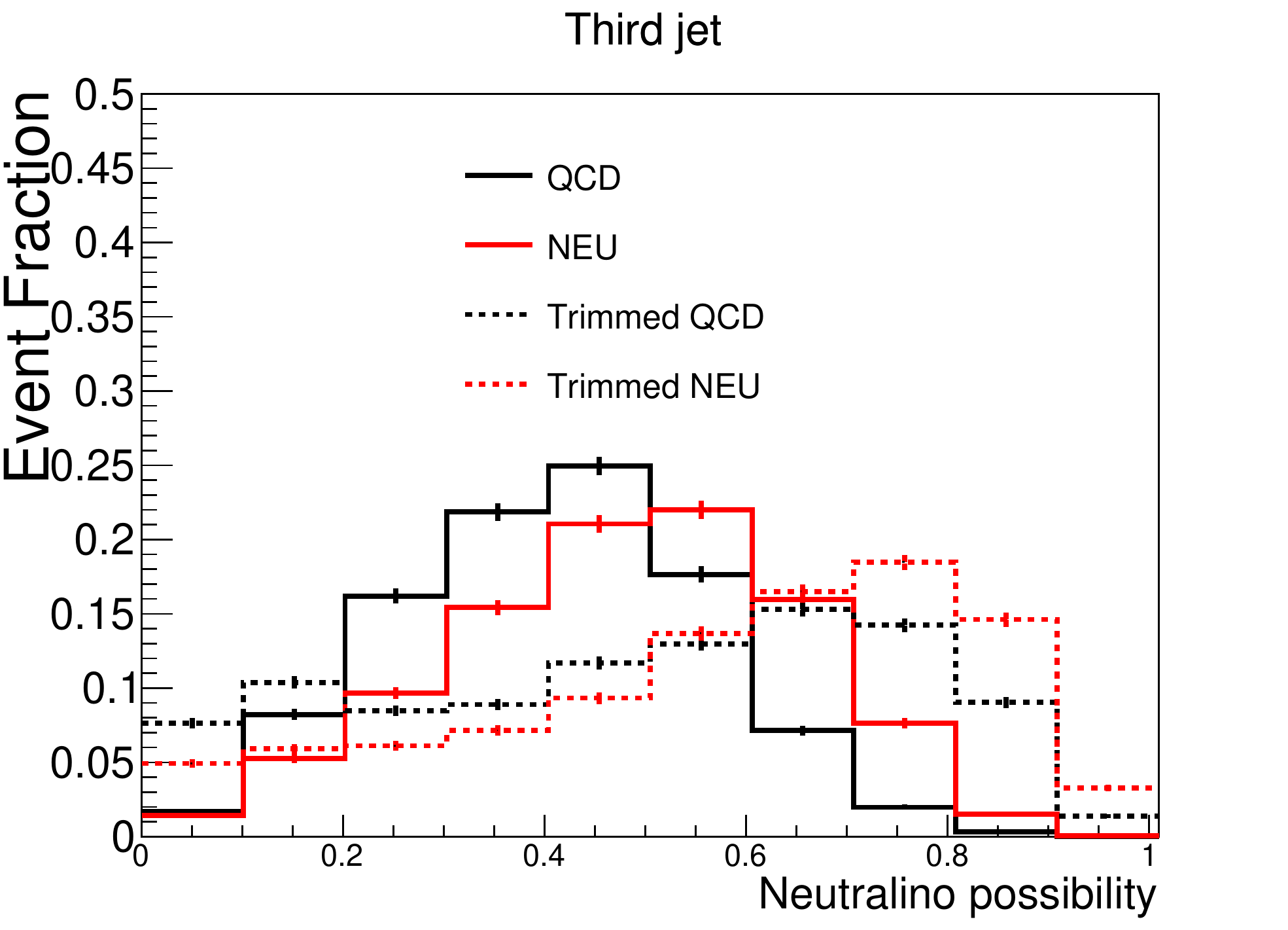}
\end{center}
\caption{\label{fig:sco} The signal possibility of the leading three jets with (dashed line) and without (solid line) trimming procedure in the selected signal (red line) and background (black line) events. The gluino mass and neutralino mass for the signal process are taken to be $1.5$ TeV and 100 GeV for illustration. }
\end{figure}

Now, we can apply the CNN tag on the jets in the selected signal and background events. Firstly, in each of the selected events, jets are reconstructed in the same way as the training sample, {\it i.e.}, anti-$k_t$ with radius parameter $R=1.0$ and transverse momentum $p_T>100$ GeV. 
Since two neutralino jets can be either energetic or relatively soft in the signal process, all the reconstructed jets are passed to our CNN for neutralino tagging. Each of them will be assigned a signal possibility (There are two outputs of the CNN: signal and background possibilities.  The background possibility is correlated with the signal possibility).
Then, jets are ranked by the signal possibility. 
The distributions of signal possibilities for the leading three jets are shown in Fig.~\ref{fig:sco}, where the gluino mass and neutralino mass are set to 1.5 TeV and 100 GeV, respectively. We can see that the jets in the signal events obtain larger signal possibility than those in background events. These information can help to separate the signal and background further. 
On the other hand, it can be readily seen from the figure that even the background jets can obtain relatively high CNN scores of signal possibility. This indicates that the neutralino jet and QCD jet in the full signal and background events after the selections are much more difficult to discriminate than those in the training sample. The difficulty is mainly attributed to the severe contaminations among jets in the selected events. 
As will be demonstrated later, those contaminations tend to make multi-prong QCD jet, which also reduces the discriminating power of the N-subjettiness variable. 
The dashed lines in the same figure, correspond to the CNN tagging efficiencies on jets after preforming the jet trimming, with trimming parameters the same as in the ATLAS analysis~\cite{ATLAS-CONF-2016-057}. 
Due to the hardness of the contamination, the trimming fails to resolve the jets. 

\begin{figure}[htb]
\begin{center}
\includegraphics[width=0.48\textwidth]{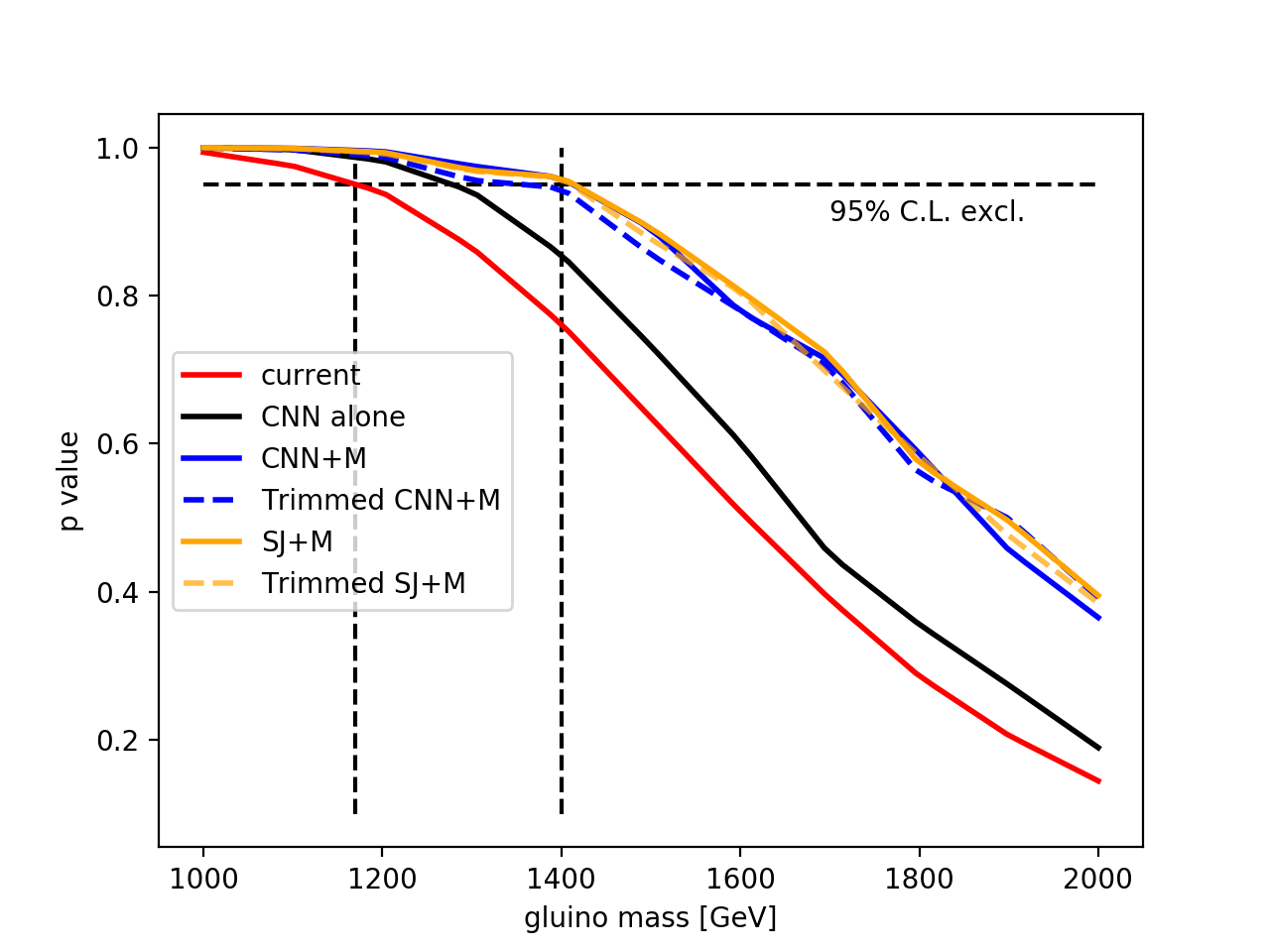}
\includegraphics[width=0.48\textwidth]{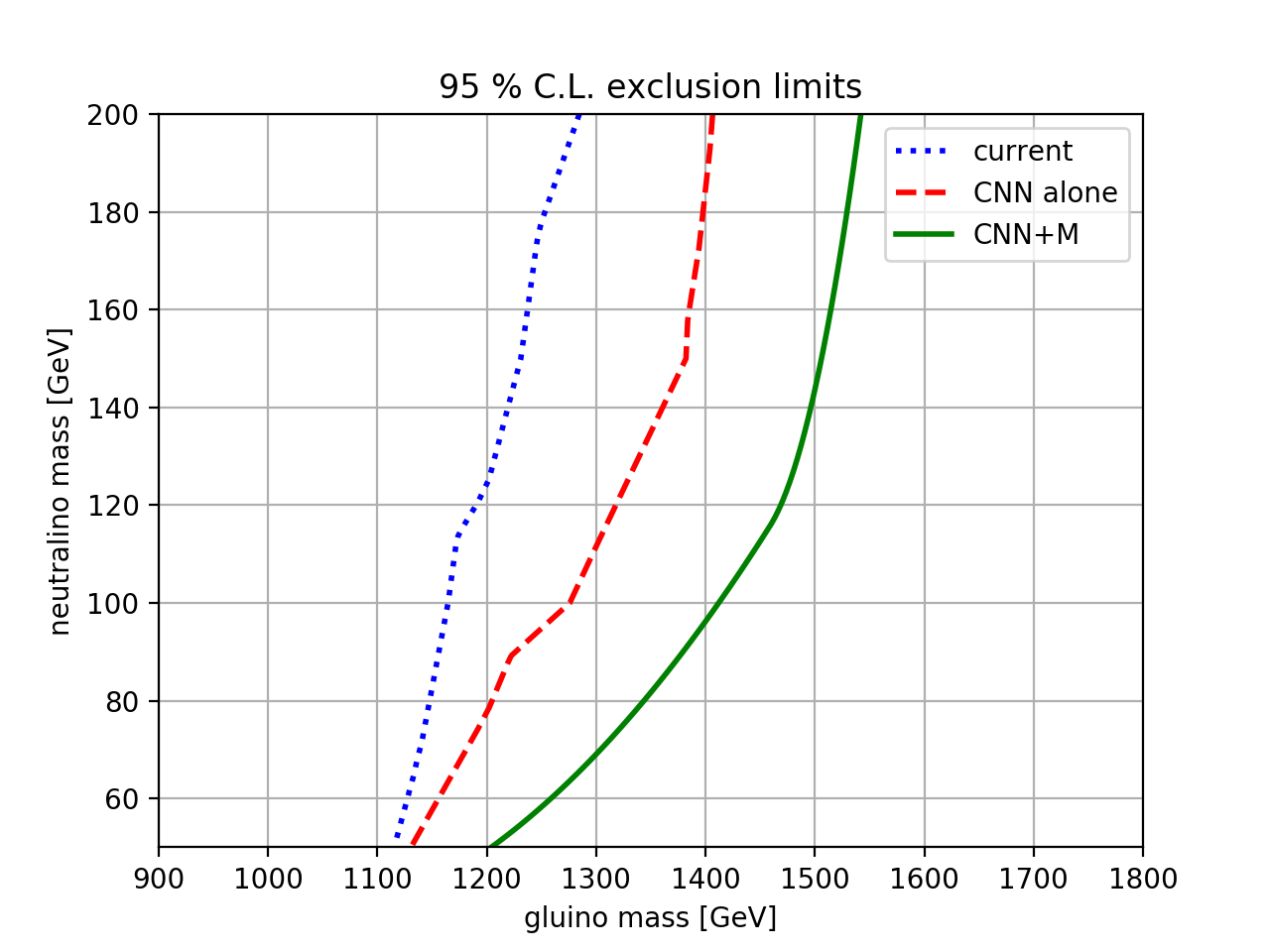}
\end{center}
\caption{\label{fig:bound} Left: The $p$-values for the original ATLAS analysis (red solid line) and BDT analyses with either CNN alone (dark solid line) or with various additional information as indicated in the legend. Right: The 95\% C.L. exclusion limits for the original ATLAS analysis (blue dotted line), CNN alone analysis (red dashed line) and CNN+M analysis (green solid line). }
\end{figure}

Because the CNN scores (signal possibility) for the leading three jets of signal and background are correlated to some extent. We employ again the BDT method to study the discriminating power of the combination of these informations (including jet invariant mass and N-subjettiness). 
Compared to the BDT analysis in the previous section, we now have much fewer available events ($\sim 10$K) and more input variables (CNN scores and invariant masses of leading three jets). However, the same BDT parameters turn out to perform quite well here. At each mass point, this BDT is trained on 5000 signal and 5000 background events, and it is tested on the rest independent events. The Kolmogorov-Smirnov tests are found to be always greater than $\sim 0.1$ for both signal and background, which indicate that the BDT is free from the over-trainning problem. 
For each BDT trained and validated on given gluino and neutralino masses, applying a cut on its BDT response will reduce the signal and background cross section further down to $\sigma^{13}(\tilde{g}\tilde{g}) \times \epsilon^{\text{4jSRb1}} \times \epsilon^{\text{BDT}}_S$ and $\sigma^{\text{BG}} \times \epsilon^{\text{BDT}}_B$, respectively. The $\epsilon^{\text{BDT}}_{S/B}$ corresponds to the selection efficiency of a BDT cut on signal/background events. 
We assume the observed event number is reduced by the same factor of $\epsilon^{\text{BDT}}_B$. 
As for the background uncertainties, the statistical component is rescaled by a factor of $\sqrt{\epsilon^{\text{BDT}}_B}$ while others are rescaled by the factor of $\epsilon^{\text{BDT}}_B$.
We will adopt the $p$-value (or CLs)~\cite{Read:2002hq} to characterize the probability of the signal exclusion, which is defined as 
\begin{align}
p=1- \frac{\mathcal{P} (\mathcal{H}_{S+B})}{\mathcal{P} (\mathcal{H}_{B})},
\end{align}
where $\mathcal{P} (\mathcal{H}_{S+B})$ and $\mathcal{P} (\mathcal{H}_{B})$ are probabilities of signal plus background hypothesis and background only hypothesis, respectively. 
So the $\mathcal{H}_{S+B}$ hypothesis is excluded at 95\% C.L. if the $p$-value is greater than 0.95. 
The BDT cut that maximizes the $p$-value will be taken at each gluino-neutralino mass point in each analysis~\footnote{The optimized BDT cut efficiencies are found to be $\sim$50\% for signal and 10-20\% for background. So the signal and background event numbers in our simulation after the BDT cut is around 5000 and 1000-2000, respectively.}. 
The $p$-values for the original ATLAS analysis and our BDT analyses with either CNN output alone or with combined information are shown in the left panel of Fig.~\ref{fig:bound}, where we have fixed $m_{\tilde{\chi}^0_1}=100$ GeV.
Our recasting of the ATLAS analysis shows that the benchmark points with gluino mass below $\sim 1.18$ TeV can be excluded at 95\% C.L, which coincides with the experimental result. 
Including the information of the CNN output alone will push the lower bound of the gluino mass to $\sim 1.3$ TeV. 
By adding the jet invariant mass into BDT, the non-observation of any excess will exclude the gluino mass lighter than $\sim 1.4$ TeV. 
For comparison, we have also shown the $p$-values for the analysis with the information of N-subjettiness and jet invariant mass, which does not perform better than the CNN+M analysis. Furthermore, the dashed lines correspond to the $p$-values of the analyses in which the jets are trimmed before performing the tagging. It turns out the trimming procedure does not help improving the signal and background discrimination. 

We have demonstrated that the CNN+M method (without trimming) provides one of the most sensitive probes for the RPV gluino search.  
Finally, we show the application of the method (with the CNN trained on the $m_{\tilde{\chi}^0_1}=100$ GeV events sample) to 
the two-dimensional $m_{\tilde{g}}$-$m_{\tilde{\chi}^0_1}$ plane. 
In the right panel of Fig.~\ref{fig:bound}, the 95\% C.L. exclusion limits for the original ATLAS analysis, the CNN alone analysis and CNN+M analysis are given. Here the CNN is trained on the event sample with $m_{\tilde{\chi}^0_1} =100$ GeV. We can observe that such a CNN is vulnerable to lower neutralino mass, {\it i.e.}, the improvement is dramatically decreased for the neutralino mass less than 100 GeV, while it is much less sensitive to the higher neutralino mass. This is mainly because the neutralino with mass $m_{\tilde{\chi}^0_1} \lesssim 200$ GeV from heavy gluino decay has transverse momentum larger than $\sim 400$ GeV. All its decay products are captured by the jet reconstruction. So the jet substructure is detectable except when the neutralino is 
so light that its subjets become overlapping. 
Including the jet invariant mass information can help pushing the gluino bounds by $\sim 100$ GeV higher. In particular, the jet invariant mass has better discriminating power for heavier neutralino mass, compensating for the slight decrement of neutralino jet tagging efficiency, which can be also seen from Fig.~\ref{fig:masses}.

\section{conclusion} \label{sec:conc}

In the paper, we study the possible improvement on current hadronic RPV search by BDT method with  information from jet substructure. In particular, the convolutional neutral network is adopted to tag the neutralino jet which decays into three quarks. The application of the CNN to an existing RPV gluino search by the ATLAS Collaboration in final state with multiple energetic jets is investigated. 

The information of jet can be formatted into jet image by identifying each calorimeter cell as a pixel. The energy distribution of all particles, the energy distribution of charged particles and the number of charged particles in calorimeter cells are regarded as the RGB color of those pixels. 
The CNN is trained on events of simplified process with only a visible neutralino jet in the final state. 
So it is able to tag a neutralino jet by using the jet image irrespective of its production mechanism.
According to the small size and sparsity of jet image, the \texttt{VGGNet} CNN architecture, 
which was optimized for \texttt{CIFAR-10} dataset, has been adopted.  
It is able to tag the neutralino jet with efficiency of 50\% while only accept $\sim $1\% of QCD jet. These efficiencies were found to be insensitive to the CNN parameters in a wide range. 
Moreover, due to the cylinder shape of detector,  jet image has strong dependence on the pseudorapidity of the jet. The CNN performs well for jet either in the central region ($|\eta|\to 0$) or with relatively large pseudorapidity ($|\eta| \lesssim 2.5$). 
Our CNN can outperform the high-level jet substructure variable N-subjettiness by a factor of a few in neutralino jet and QCD jet discrimination.  
However, the jet invariant mass information is not fully learned by the CNN, partly because the image preprocessing 
does not respect the Lorentz symmetry. Combining the CNN output with the jet invariant mass can improve the signal efficiency further. More importantly, for the CNN being trained on a given neutralino mass, the CNN+M tagging method performs much better than the method with the CNN alone when applied to the neighbor of that neutralino mass.

To study the realistic application of the CNN, the ATLAS analysis is recasted. Only the events (for both signal and background)
which can pass all selection cuts of the 4jSRb1 signal region in the ATLAS analysis were kept. The CNN assigns 
``neutralino jet possibilities'' to all jets in these events. The jets in signal events are likely to obtain higher 
``neutralino jet possibilities'' than those in background events. 
Compared to the simplified processes (for generating training sample) with single target jet in the final state, the heavy contaminations due to multiple energetic jets in the final state greatly reduce the discriminating power of 
both the CNN and N-subjettiness. But the BDT analysis with information from the CNN scores of three leading jets is still able to push the lower bound of the gluino mass by $\sim 100$ GeV. 
The combined analyses of either CNN+M or N-subjettiness+M have similar sensitivities, 
{\it i.e.}, excluding the gluino mass lighter than $\sim$1.4 TeV for $m_{\tilde{\chi}^0_1} = 100$ GeV. 
By applying the CNN and CNN+M analyses to the two-dimensional $m_{\tilde{g}}$-$m_{\tilde{\chi}^0_1}$ plane, 
we found the CNN tagging efficiency is vulnerable to lighter neutralino while is insensitive to heavier neutralino up to $\sim 200$ GeV. The CNN+M method can help pushing the gluino bounds by 100-250 GeV higher depending on the neutralino mass.

\section*{Acknowledgement}
We thank the Korea Institute for Advanced Study for providing computing resources (KIAS Center for Advanced Computation Linux Cluster System) for this work.
This research was supported in part by the Projects 11475238 and 11747601 supported
by National Natural Science Foundation of China, and by Key Research Program of Frontier Science, CAS (TL)
and by National Research Foundation of Korea (NRF) Research Grant NRF-2015R1A2A1A05001869 (JL). 

\phantomsection
\addcontentsline{toc}{section}{References}
\bibliographystyle{jhep}
\bibliography{DLrpv}

\end{document}